# Influence of Ni/Mn cation order on the spin-phonon coupling in multifunctional La$_2$NiMnO$_6$ epitaxial films by polarized Raman spectroscopy

**K.D. Truong, M.P. Singh \*, S. Jandl, and P. Fournier**

*Regroupement québécois sur les matériaux de pointe,
Département de physique, Université de Sherbrooke
Sherbrooke, Québec J1K 2R1, Canada*

## Abstract

We report the influence of Ni/Mn ordering on the spin-phonon coupling in multifunctional La$_2$NiMnO$_6$. Three types of films with different levels of structural order, including long-range Ni/Mn cation order, cation disorder, and an admixture of the ordered and disordered phases, are compared by polarized micro-Raman spectroscopy and magnetometry. Each film displays a strong dependence on the polarization configuration and a unique set of Raman active phonon excitations. Long-range cation ordering results in the splitting of Raman active phonon peaks because of Brillouin zone folding and lowering symmetry. Phonon mode softening begins clearly at a distinct temperature for each sample revealing a strong spin-lattice interaction. It follows closely the magnetization curve in ordered films. Unlike the admixture and the ordered films, softening behavior is strongly suppressed in the cation-disordered films. These differences may be understood based on the variation in amplitude of the spin-spin correlation functions due to the local Ni/Mn cation ordering.

\* Corresponding author (email: mangala.singh@usherbrooke.ca)

## I. Introduction

Double perovskite La$_2$NiMnO$_6$ (LNMO) is a polar ferromagnetic insulator. Consequently, it has received considerable interest owing to its multiferroic behaviour [1] that can be promoted by manipulating the coupling between its electronic, magnetic and phonon order parameters [2]. Several studies have focused on the structural, electronic, and magnetic properties of LNMO [2-16]. They have undoubtedly revealed that the magnetic properties of LNMO are very sensitive to the local Ni/Mn cation ordering [2-4, 6-16]. As shown with bulk samples, ordered LNMO should display a lone paramagnetic-to-ferromagnetic transition [2] at about 300 K while samples composed of both ordered and disordered phases should have two magnetic phase transitions [3] at ~ 150 K and ~ 300 K arising from distinct magnetic phases. Thus, the magnetic behavior of LNMO can be used to establish the Ni/Mn cation ordering in LNMO. As was found also in bulk materials [12], the ordered phase possesses alternating Ni$^{2+}$ ($t_{2g}^6 e_g^2$) and Mn$^{4+}$ ($t_{2g}^3 e_g^0$) planes leading to a polar system with large dielectric constant, while such effect is absent in a disordered phase where both Ni and Mn cations are randomly distributed in the crystal structure and both cations have the 3+ oxidation states [6-16]. High temperature ferromagnetic transition in LNMO arises from the Ni$^{2+}$-O-Mn$^{4+}$ superexchange interaction while the low temperature magnetic transition arises due to the Ni$^{3+}$-O-Mn$^{3+}$ superexchange interaction. Various factors ranging from oxygen/cations non-stoichiometry and synthesis/processing conditions are expected to provoke a disordered structure in these double perovskites [2-16].

It is well known that the functional properties of strongly correlated oxides are typically governed by the competitive interactions between spin, phonon, and polarization order parameters. In this perspective, local cation ordering in double perovskites may play a crucial role in determining the multiferroic coupling [1] and functional properties. Indeed, it has been recently observed that the magnetodielectric effect in long-range ordered bulk LNMO is only present in the vicinity of a *ferromagnetic-to-paramagnetic* transition [2] and a similar behaviour was also demonstrated for the long-range ordered La$_2$CoMnO$_6$ (LCMO) films [17]. On the contrary, there is no such coincidence in short-range ordered LNMO thin films [18] as the large magnetodielectric response shows up around 150 K far away from its ferromagnetic-to-paramagnetic transition temperature at 280 K observed by magnetization [5]. This illustrates the likely influence of local cation ordering on the spin-polarization coupling.

Recently Iliev et al. [18] have shown a strong spin-phonon coupling and a short-range ordering in LNMO films using polarized Raman spectroscopy [19]. To our knowledge unlike the LCMO [20, 21], no systematic study has illustrated the likely impact of Ni/Mn cations ordering on the spin–phonon interactions in LNMO. Such studies are of crucial importance to understand the underlying physics behind the multifunctional coupled behaviours in double perovskites. In this work, we study three types of epitaxial LNMO films consisting of: 1) a long-range cation-ordered phase; 2) a cation-disordered phase with random positioning of the Ni/Mn cations; and 3) an admixture phase (called also short-range ordered). We report the relation between the Raman active phonons and the level of ordering in the films as we demonstrate the gradual evolution of the corresponding phonon spectrum. Furthermore, we explore the temperature dependence of the Raman scattering response for various polarization configurations and show a clear relationship between the Raman mode softening and the ferromagnetic-to-paramagnetic transitions. The roles played by the disordered and the ordered structures in the spin-phonon coupling of LNMO are discussed and compared with previous studies.

## II. Growth details

Epitaxial thin films of LNMO on $SrTiO_3$ (111) were grown under $O_2$ ambient at pressure ranging from 100-1000 mTorr using a pulsed laser deposition system. Growth temperature from 475-850 $^o$C range was explored. A stoichiometric polycrystalline LNMO target was synthesised by the standard solid-state synthesis method and its cationic stoichiometry was subsequently verified using an energy dispersive X-ray spectroscopy system equipped with a scanning electron microscope. The films composed of the long-range ordered phase were grown at 800 $^o$C/800 mTorr $O_2$, the ones with the disordered Ni/Mn phase at 500 $^o$C /300 mTorr $O_2$ while the admixture of both the ordered and the disordered phases was obtained at 800 $^o$C /300 mTorr $O_2$. Following the deposition, films were cooled down to room temperature with 400 Torr $O_2$ at a rate of 10 $^o$C/min. These films were not subjected to any post-deposition annealing process. Influence of the growth conditions on the structural properties and the phase formation of LNMO thin films and the detailed phase-stability diagram is published elsewhere [22]. The presence of the superlattice reflections in the X-ray diffraction patterns of our films demonstrates Ni/Mn long-range ordering and emphasizes the necessity to use growth conditions in a critical region of the pressure-temperature phase diagram to obtain a controlled Ni/Mn ordering in LNMO [22]. The dependence of cation ordering on the growth parameter is also consistent with the phase-stability diagram of PLD-grown $La_2CoMnO_6$ films [17]. It shows that both ordered and disordered phases can be stabilized only within very narrow growth parameters while the admixture phase can be obtained in a wide range of growth parameters with varying proportions of the ordered and disordered phases [22]. We selected the admixture film for this work with respect to many other films grown in other conditions using the temperature dependence of the magnetization which shows clearly two distinct transitions [3] as signatures of the long-range ordered and the disordered phases (see below).

### III. Results and Discussion
### A. Magnetic properties

In order to examine the Ni/Mn cation ordering in our films and to establish a correlation between the magnetic properties and the behaviour of the Raman active phonons, we first measured the temperature (T) dependence of the magnetization (M) curves (*i.e.,* M-T curves) and its magnetic field (H) dependence (*i.e.,* M-H loops) with a Superconducting quantum interference device (SQUID) magnetometer from Quantum Design. The M-T curves (Fig. 1a) show that films grown under 500°C/300mTorr $O_2$ exhibit a single ferromagnetic-to-paramagnetic transition at ~138 K illustrating that the films possess a random distribution of $Ni^{3+}/Mn^{3+}$ cations [6-16]. By contrast, the 800°C/800 mTorr $O_2$ films exhibits also a single ferromagnetic-to-paramagnetic transition at 275 K [6-16]. This indicates that these latter films contain a long-range ordered $Ni^{2+}/Mn^{4+}$ cation sequence similar to that observed in the bulk [2]. Films grown at 800 °C /300 mTorr are characterized by two magnetic transition temperatures at ~ 140 K and ~ 295 K which indicates clearly that they contain both the long-range ordered and the disordered phases [2-16]. For this admixture film, we can estimate roughly to approximately 30% the proportion of the disordered phase using the magnetic behavior based on the magnetic properties of the long-range ordered and disordered films. The typical M-H loop recorded on well-ordered films is shown in the inset of Fig.1. Long-range ordered films are characterized by a saturation magnetization of roughly 4.8 $\mu_B$/f.u. which is close to the expected value of 5$\mu_B$/f.u. The magnetic properties of these films, thus, clearly establish the global nature of Ni/Mn cation ordering in our samples. It is here important to note that magnetic properties of the films provide a qualitative information about the Ni/Mn cation ordering at the B-sites. Further study may be therefore warranted to quantify the Ni/Mn ordering in these films using neutron and soft x-ray scattering techniques.

The high temperature ferromagnetic transition in LNMO is governed by the 180° $Ni^{2+}$-O-$Mn^{4+}$ superexchange process as mentioned above and therefore one might expect the same value of FM-$T_c$ in both the admixture and the long-range ordered films. Strikingly, the value of 295 K for the FM-$T_c$ in the admixture LNMO films is appreciably larger than that of the well-ordered LNMO films with FM-$T_c$ ~ 275 K. This difference could be understood as follows. The value of the magnetic Curie temperature due to the superexchange process is determined by the magnitude of the spin-transfer integral which depends on the degree of Ni/Mn-O orbital overlap and thus, it increases exponentially with decreasing $Ni^{2+}$-O-$Mn^{4+}$ bond length. The in-built local electric field originating from the alternative stacking of $Ni^{2+}$/$Mn^{4+}$ cations at B-sites in well-ordered LNMO enhances its Coulombic energy. This provokes an expansion in lattice parameter owing to a net enhancement in $Ni^{2+}$-O-$Mn^{4+}$ bond lengths helping to minimize the elastic energy [22]. On the contrary, elastic energy minimization in short-range ordered LNMO is partially aided by the presence of the disordered phase in our admixture films and therefore a relatively smaller $Ni^{2+}$-O-$Mn^{4+}$ average bond length resulting in a larger superexchange strength and hence the value of FM-$T_c$. This proposed change in $Ni^{2+}$-O-$Mn^{4+}$ bond length in LNMO owing to structural ordering is expected to have an observable effect on the Raman active phonon behaviors.

### B. Raman Data

Using magnetic properties of these films, we established a first impact of Ni/Mn ordering in our films. Polarized Raman spectroscopy is a very sensitive technique to investigate the local and/or dynamical structural properties, charge/orbital ordering, and spin-phonon coupling in strongly correlated perovskite oxides [23, 24]. It has been successfully extended to study the impact of cation ordering on the Raman phonon behavior and spin-phonon coupling in $La_2CoMnO_6$, [21] short-range ordered LNMO films [18, 19] and disordered $LaBiCo_{1/4}Mn_{3/4}O_6$ films [25]. We therefore used it to explore the influence of local Ni/Mn cation ordering on the spin-phonon coupling and Raman phonon behavior in LNMO. Raman spectra of our LNMO films were measured in the 10-300 K temperature range under different polarization configurations (*viz.,* X'X', XX, XY, X'Y'), where X and Y refer to the parallel rectangular edges of the STO (111) substrate (one is along the (1-10) direction) while X' and Y' refer to its diagonal axis as shown in the inset of Fig 1b. [18] A

Labram-800 microscope spectrometer equipped with a He-Ne laser and a nitrogen-cooled charge coupled device (CCD) detector was used for the measurement while the film is mounted on the cold finger of a Janis Research Supertran cryostat. To avoid the heating of our films during the measurements, power of the laser beam was limited to 0.3mW through a 50x objective focusing about the excitation light on a 3 µm diameter-sized spot.

The total number of allowed Raman excitations and their dependence on the polarization configuration for LCMO has been recently computed using the lattice dynamical model by Iliev et.al. (Ref. 20) and has been successfully used in our previous study of LCMO's Raman spectra [21]. Since both LNMO and LCMO exhibit a similar impact of Ni (Co) and Mn ordering on the crystal symmetry and the lattice parameters[12], theoretical predictions for LCMO double perovskite [20] can safely be extended to identify the Raman modes observed in LNMO samples. Indeed, it has recently been used to understand successfully the properties of LNMO films as well as single crystals [18]. We have therefore adopted these theoretical results to interpret and to elucidate the polarization dependence Raman spectra of our films. A detailed discussion on this matter can be found in Refs. 18 and 20.

Typical polarization dependences of Raman active phonon spectra of our ordered LNMO films at room temperature are shown in Figure 1b. The intensity of the phonon excitations is strongly dependent on the polarization configurations, namely XX, XY, X'Y', and X'X' as a result of the epitaxial character of our films. As pointed out in Refs. 18 and 20, strong modes observed around 668 and 527 $cm^{-1}$ can be assigned to stretching (S) and anti-stretching (AS) vibrations of the (Ni/Mn)$O_6$ octahedra, respectively. For the 527 $cm^{-1}$ AS mode, $O^{2-}$ ions vibrate perpendicularly to the Ni/Mn-O bond whereas for the 668 $cm^{-1}$ S mode, Ni/Mn-O bond involves both in-plane and out-of-plane vibrations. In the monoclinic $P2_1/n$ symmetry, only 24 (12$A_g$+12$B_g$) modes are Raman active. Presence of a weak stretching mode in the X'Y' spectra indicate that the films possess a monoclinic $P2_1/n$ symmetry, which is well-consistent to its structural properties [22]. When the samples are cooled down, drastic enhancements in intensities of both symmetric and anti-symmetric modes (Figs. 2-5) are observed at low temperature irrespective of Ni/Mn ordering in the films.

However, the total number of Raman active excitations and their spectral characteristics differ significantly from one type of samples to the other. To understand the differences and/or similarities arising from Ni/Mn cation ordering, we present in Figure 2 typical Raman spectra recorded in the XX and XY configurations for these samples at 10 K. In the XX configuration, disordered films are characterized by broad stretching and AS modes with relatively low intensity while these modes evolve as sharp peaks with relatively large intensity in the admixture and the ordered films. For example, 540 cm$^{-1}$ AS in XX configuration (shown as arrow in Fig 2a) mode in the disordered film evolves into a well-defined peak in admixture film (529 cm$^{-1}$) whereas two peaks (viz., 503 cm$^{-1}$ and 527 cm$^{-1}$) are observed within its closed proximity in the ordered films. A similar progressive trend can also be observed in the symmetric 668 cm$^{-1}$ Raman excitation (Fig 2a). As the Ni/Mn ordering diminishes, it becomes a very broad asymmetric peak (marked as-arrow in Fig 2a) with relatively low intensity in the disordered phase. In XY configuration, a similar comparative distinction between the characteristics of various Raman excitations as a function of Ni/Mn ordering is easily noticeable. More importantly, ordered films exhibit additional Raman active phonons irrespective of polarization configurations. For example, in the XY configuration, the stretching modes of the disordered films are characterized by a broad peak which includes in fact two modes at 643 cm$^{-1}$ and 666 cm$^{-1}$. As Ni/Mn ordering improves, it evolves into two clear Raman modes at 648 cm$^{-1}$ and 665 cm$^{-1}$ in the admixture phase (marked by arrows), and about four well-defined active Raman modes at 621 cm$^{-1}$, 630 cm$^{-1}$, 645 cm$^{-1}$, 665 cm$^{-1}$ in the long-range ordered phase. These same Raman active modes in the disordered films have relatively large full-width at half maximum (FWHM ~ 33 cm$^{-1}$) compared to the ordered film (FWHM ~ 23 cm$^{-1}$) at 10 K. It can be interpreted as a direct consequence of the random distribution of the Ni and Mn cations in the disordered films. A similar trend in the increase of the total number of Raman allowed excitations can be also observed in the XX configuration.

The other important feature is the presence of weak intensity modes (insets of Fig. 2a and 2b) in the low frequency regime. One may clearly note a progressive enhancement in their intensity and their total number. The emergence of these additional phonon excitations owing to the presence of a secondary impurity phase in our films can be easily ruled out based on their temperature dependence magnetic properties (Fig. 1a) [2,3,7,16,17]. As we mentioned earlier, a presence of potential impurity phase in LNMO and as-well-as in similar double perovskites induces additional low-temperature magnetic transitions [3]. The presence of these low-frequency Raman excitations in the ordered phase and their significant reduction in the admixture and finally their absence in the disordered films reveals that these excitations do not also arise due to the density of states owing to the Ni/Mn disorder. In fact, this trend reveals that these bands arise due to the long-range Ni/Mn ordering. These spectra are also well-consistent with Raman spectra of LNMO single crystals[18]. In Table 1, we summarize the observed phonon excitations for the three types of samples. Our spectral analysis at 10 K shows that the ordered films display a large number (viz., 12 excitations) of phonon excitations compared to the disordered (viz., 7 excitations) and short-range ordered films (viz., 7 excitations), which is also well-consistent with Iliev *et al.* predictions [18, 20] and our previous work [21] on LCMO. Thus, a clear progression in the increase of the total number of Raman active modes, the peak broadening (sharpness), and their spectral characteristics is observed by moving from a disordered to the ordered films.

The origin of Raman active excitations in the disordered films can be understood as follows. It is here important to note that an average ideal-cubic perovskite unit cell does not exhibit any Raman active modes. A deviation from an ideal cubic symmetry to the lower crystal symmetry is however expected in $LaNi_{0.5}Mn_{0.3}O_3$ owing to the rotation of (Mn/Ni)$O_6$ octahedra and the difference in the size of various ions in the unit cell. The global crystal symmetry of an $ABO_3$ type perovskite can be determined by the tolerance factor "t", which is defined as $r_A+r_O/1.41(r_B+r_O)$, where $r_A$, $r_O$, $r_B$ are the respective ionic radii of the A, B and O ions [13]. For t = 1, the system ideally exhibits a cubic symmetry, for vale of "t" in the range of 1 to 0.95, it possesses tetragonal symmetry, while for the value of "t" in the range of 0.95 to 0.8, it either possesses the orthorhombic or monoclinic symmetry and above 1, it possesses the hexagonal symmetry. Similarly, the instability in the lattice of doped perovskite $AB'_{0.5}B''_{0.5}O_3$ or $A_2B'B''O_6$ (A remains the rare earth atom while B' and B'' are transition metal ions) from ideal cubic symmetry can be determined by the tolerance factor $t = (r_A + r_O)/1.41(<r_B> + r_O)$, where $r_A$, $<r_B>$ and $r_O$ are the on-site ionic radius of A, average B'/B'', and O ions, respectively. For $LaNi_{0.5}Mn_{0.5}O_3$ and $La_2NiMnO_6$, the value of "t" is about 0.85 and 0.84, respectively leading to either an orthorhombic or a monoclinic symmetry. Apart from tolerance factor, oxidation states of Ni/Mn cations, crystal fields, and local polarizations ultimately determine the crystal structure of $LaNi_{0.5}Mn_{0.5}O_3$ and $La_2NiMnO_6$. The experimental studies on bulk materials demonstrate that an ordered LNMO system [2, 13] exhibits the monoclinic $P2_1/n$ symmetry (a = 5.52 Å, b=7.74 Å, c= 5.46 Å, and β = 90.04 °) while the disordered system possesses the orthorhombic Pnma symmetry (a = 5.50 Å, b= 5.54 Å, c= 7.73 Å, and β =90°). Recent studies on LNMO thin films clearly demonstrate that the admixture films comprise structural domains [3-4] of having both orthorhombic and monoclinic symmetry with varying sizes. Thus, the presence of Raman active modes in the disordered films is possible to assign to the orthorhombic symmetry of $LaNi_{0.5}Mn_{0.5}O_3$ [20].

Compared to the disordered films, the observation of additional Raman phonons in the ordered phase can be understood as follows. In an ideal disordered perovskite sample, Ni and Mn are randomly distributed at B-sites (Fig. 6) in an ideal primitive pseudo-cubic $ABO_3$ (Fig. 6) perovskite unit cell. However, the Ni and Mn are alternatively arranged at the B-sites in a long-range ordered double perovskite. The resulting pseudo-cubic unit cell (Fig. 6) then shows increased lattice parameters ( ~ $2a_P$) with respect to the primitive pseudo-cubic $ABO_3$ unit cell which eventually causes Brillouin-zone folding and the observation of new Γ-point active Raman excitations. In the admixture phase, nanosized domains of the ordered phase coexist with disordered domains, [3, 4] and a Raman spectrum measured in a specific polarization configuration reflects an average of the separate contributions of Brillouin zone folding and disorder (Fig. 2). Additionally, the increase in the number of distinct phonon excitations (Table I) from the disordered and the admixture to the ordered films is further aided by a significant enhancement in the phonon lifetime. The observed Raman active modes in the disordered films are governed by the crystal symmetry of $LaNi_{0.5}Mn_{0.5}O_3$ and not due to the density of states owing to the random distribution of Ni/Mn as pointed out earlier. Our study thus clearly demonstrates that the local Ni/Mn ordering determines the observed number of Raman excitations in a given sample and the observation of additional excitations arising due to Brillouin zone folding may be considered as a generic feature to differentiate the long-range cation order in double perovskites from their disordered or their short-range ordered counterparts [18-21]. The present study may therefore be considered as a double-perovskite benchmark example of the expected behaviour of

Raman modes when these materials and other similar structures go through such a subtle structural transition.

As alluded above, we looked into the spectral characteristics of stretching mode of (Ni/Mn)$O_6$ octahedra at room temperature to understand the difference observed in value of FM-$T_c$ in admixture film (FM-$T_c$ ~ 295 K) and ordered film (FM-$T_c$ ~ 275 K). In admixture films, it arises at about 675 cm$^{-1}$ while in ordered films it arises about 672 cm$^{-1}$ (Fig. 7b and 7c) at room temperature. This illustrates that the average bond length of Ni/Mn-O is relatively longer in well-ordered films in contrast to the admixture ones. Hence, a relatively large overlap of electronic wave functions along the Ni$^{2+}$-O-Mn$^{4+}$ bond is expected in admixture films and therefore a significant enhancement in strength of superexchange interactions. This is well-consistent with the observed differences in the magnetic Curie temperature of admixture films and well-ordered films as mentioned above.

As we lower the temperature, various Raman active modes show a clear softening (Figs. 3-5). A similar softening effect was also observed in thin films and bulk of LNMO and LCMO [18-21]. In order to avoid ambiguity arising from the overlap of various Raman active modes, we followed the temperature dependence of the stretching phonon mode around 672 cm$^{-1}$ arising from (Ni/Mn)$O_6$ octahedra in the XX configuration (Figs 3a, 4a, and 5a). The temperature dependence of the Raman active mode frequency is plotted in Fig. 7 for all three types of films. Various interesting features may be noted. *First*, all films display the softening. The magnitude of the softening in the disordered films is only about 2.5 cm$^{-1}$ whereas it reaches values as large as 6 cm$^{-1}$ in the ordered films and about 7 cm$^{-1}$ in the admixture films. This illustrates that cation ordering influences drastically the phonon softening of this mode. *Second,* softening begins at distinct temperatures and its onset point is very much dependent on the film types. This is close to the magnetic transition temperature of the disordered phase (Fig. 1a) and well-consistent with the temperature dependence of its M-T curve. Softening in the ordered film begins around 275 K and follows closely the temperature dependence of the magnetization behaviour. Similar softening effects have also been observed in short-ranged and long-ranged ordered LCMO as well as short-ranged ordered LNMO [18-21]. In the case of admixture films, we have been unable to measure the Raman spectra above 300 K owing to the temperature limitations of our Raman spectrometer. Thus, it is not possible to identify unambiguously the softening onset point for the admixture films. Nevertheless, a continuous softening from 300 K can be easily observed in the admixture films as we lower the temperature. Thus, these films show a clear softening below the magnetic transition temperature owing to the spin-phonon coupling. This presents a first glimpse of magnetoelastic coupling in our films [20].

The observed global spin-phonon coupling features and/or differences in different LNMO films could be understood as follows. The spin-phonon coupling arises from the phonon modulation of the superexchange integral [20, 21, 23-25] which depends on the net amplitude of spin-spin correlation $<S_i.S_j>$ functions where $S_i$ and $S_j$ are the localized spins at the $i^{th}$ and $j^{th}$ sites, respectively. Under the mean-field approximation, the phonon renormalization function $\delta\omega$ (T) [24] is related to the magnetization by $\delta\omega \propto M^2(T)/M_O^2$, where M(T) is the magnetization of the sample at a temperature T and $M_O$ is the magnetization at 0 K [24]. In a given sample, the amplitude of the spin-spin correlation function and hence the nature of the spin-phonon coupling will be determined by the local Ni/Mn cation order. This reveals that a random distribution of Ni/Mn cation in LNMO disordered films partially prevents softening of this mode and suppresses the spin-phonon coupling. Moreover, our study clearly manifests that spin-phonon coupling is strong in both admixture and long-range ordered films.

To summarize, we studied the impact of Ni/Mn cation order on phonon anomalies and the spin-phonon coupling in multifunctional $La_2NiMnO_6$ double perovskites. Our results display the possible correlations, similarities and/or distinctions in Raman phonon anomalies and magnetic properties caused by this Ni/Mn cation order. The long-range ordered samples show additional phonon excitations caused by Brillouin-zone folding as a clear signature of change in unit cell volume and a consequent change in crystal symmetry. The magnetic ordering at low temperature produces a softening of Raman stretching modes indicating a strong spin-lattice coupling in $La_2NiMnO_6$. These results can be easily extended to understand the possible impact of B-site ordering on the functional properties of other double perovskites which, upon long-range ordering, give rise to similar signatures and trends.

We thank S. Pelletier and M. Castonguay for their technical assistance. This work was supported by the *Canadian Institute for Advanced Research*, *Canada Foundation for Innovation*, the *Natural Sciences and Engineering Research Council* (Canada), the *Fonds Québécois pour la Recherche sur la Nature et les Technologies* (Québec) and the *Université de Sherbrooke*.

## Table caption:

**Table I :**

Lists of observed phonon excitations at 10 K in disordered, admixture and ordered films as a function of scattering XX and XY configurations.

## Figure captions

**Figure 1: (Color online) (a)** Typical M-T curves of ordered (top curve from 10 K-side), disordered (bottom curve from 10 K-side) and admixture (middle curve from 10 K-side) LNMO films. Arrows point out the magnetic transition temperatures for each type of films. The inset shows a typical M-H loop of an ordered film at 10 K. **(b)** A typical polarization configuration dependence of the Raman spectra for ordered LNMO films at 298 K. The bottom-inset shows a magnified view of the low-frequency spectral features. The top-inset shows the polarization scattering configuration used to measure the Raman spectra. Spectra have been shifted vertically for clarity.

**Figure 2: (Color online)** Raman spectra of ordered (top), admixture (middle), and disordered (bottom) films measured in the **(a)** XX and **(b)** XY polarization configurations at 10 K. Respective insets show the magnified view of the low frequency regime for the ordered (top), the admixture (middle) and the disordered (bottom) films. These spectra have been shifted vertically for clarity. Arrows show the selected excitations which evolve as a function of Ni/Mn ordering.

**Figure 3: (Color online)** Temperature dependence of the Raman spectra for an ordered film in the **(a)** XX and **(b)** XY configurations. The spectra from top to bottom are respectively measured at 10 K, 100 K, 150 K, 200 K, 250 K, and 298 K. The spectra are shifted vertically. In 3(a), the dotted vertical line at a fixed frequency is a guide to the eye to show the shift with temperature of the Raman excitation.

**Figure 4: (Color online)** Temperature dependence of the Raman spectra for a disordered film in the **(a)** XX and **(b)** XY configurations. The spectra from top to bottom are respectively measured at 10K, 100K, 150K, 200K, 250K, and 298K. The spectra are also vertically shifted. In 4(a), the dotted vertical line at a fixed frequency is a guide to the eye to show the shift with temperature of the Raman excitation.

**Figure 5: (Color online)** Temperature dependence of the Raman spectra for an admixture film in the **(a)** XX and **(b)** XY configurations. The spectra from top to bottom are respectively measured at 10K, 100K, 150K, 200K, 250K, and 298K. The spectra are shifted vertically. In 5(a), the dotted vertical line at a fixed frequency is a guide to the eye to show the shift with temperature of the Raman excitation.

**Figure 6: (Color online)** A schematic illustration of **(a)** a random distribution of B'/B'' cations; and **(b)** well-ordered B'/B'' cations at the B-sites in an ideal pseudo-cubic $ABO_3$ unit cell. Schematics of **(c)** an ideal $A_2B'B''O_6$ double perovskite unit cell; and **(d)** stacking of the $B'O_6$ and $B''O_6$ octahedra in long-range ordered $A_2B'B''O_6$. Red and yellow spheres represent B'and B''-cations respectively, big spheres represent La and small spheres represent oxygen. $B'O_6$ octahedra are represented by red and $B''O_6$ by yellow. In **(d)**, La and O atoms have been removed for clarity.

**Figure 7: (Color online)** Softening in the stretching 667 cm$^{-1}$ phonon excitation as a function of temperature for **(a)** disordered, **(b)** admixture, and **(c)** ordered LNMO films. Arrows point to the observed magnetic transition temperatures in these films. It is important to note that the highest FM-$T_c$ is close to 295 K in the admixture film which is close to the limit in temperature of our Raman spectrometer.

Table I

| Ordered | | Admixture | | Disordered | |
|---|---|---|---|---|---|
| XX | XY | XX | XY | XX | XY |
| 263 | 262 | 262 | | 262 | |
| 503 | 499 | | | | |
| 527 | 526 | 529 | 529 | 540 | 537 |
| | 621 | | | | |
| | 630 | | | | |
| 626 | 645 | 626 | 648 | 643 | 643 |
| 668 | 665 | 668 | 665 | 661 | 666 |

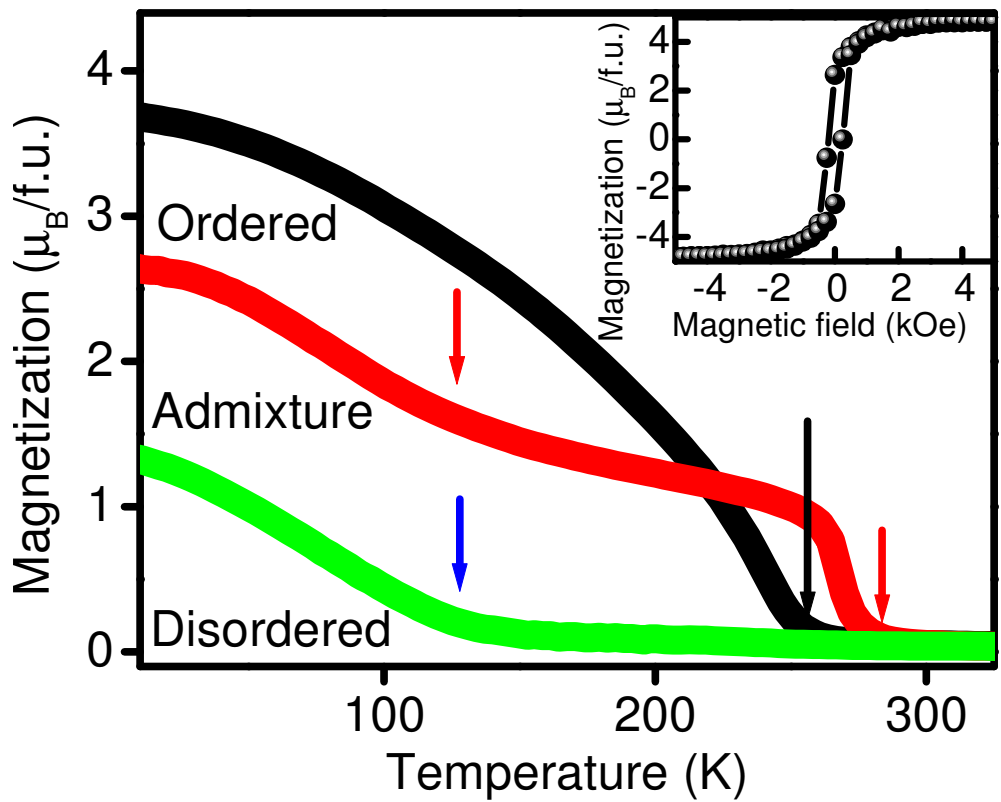
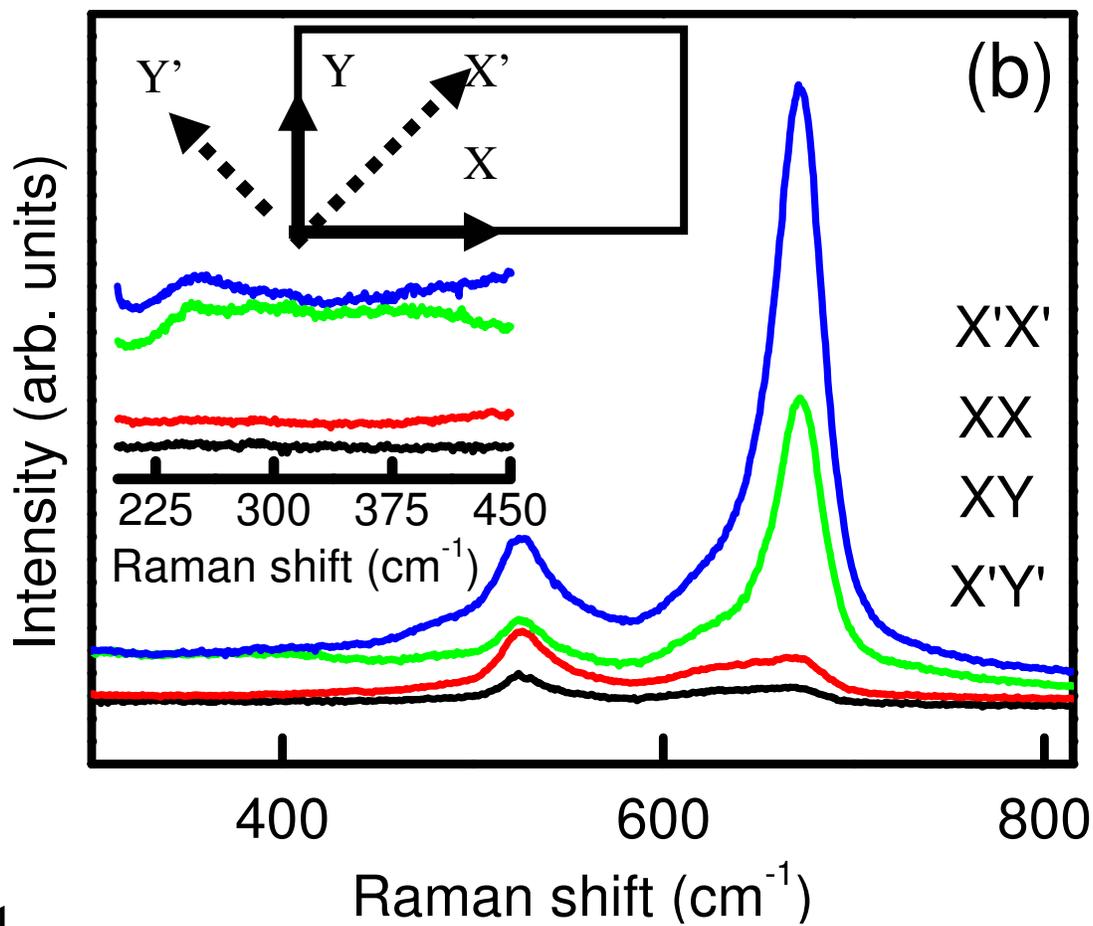

Figure 1

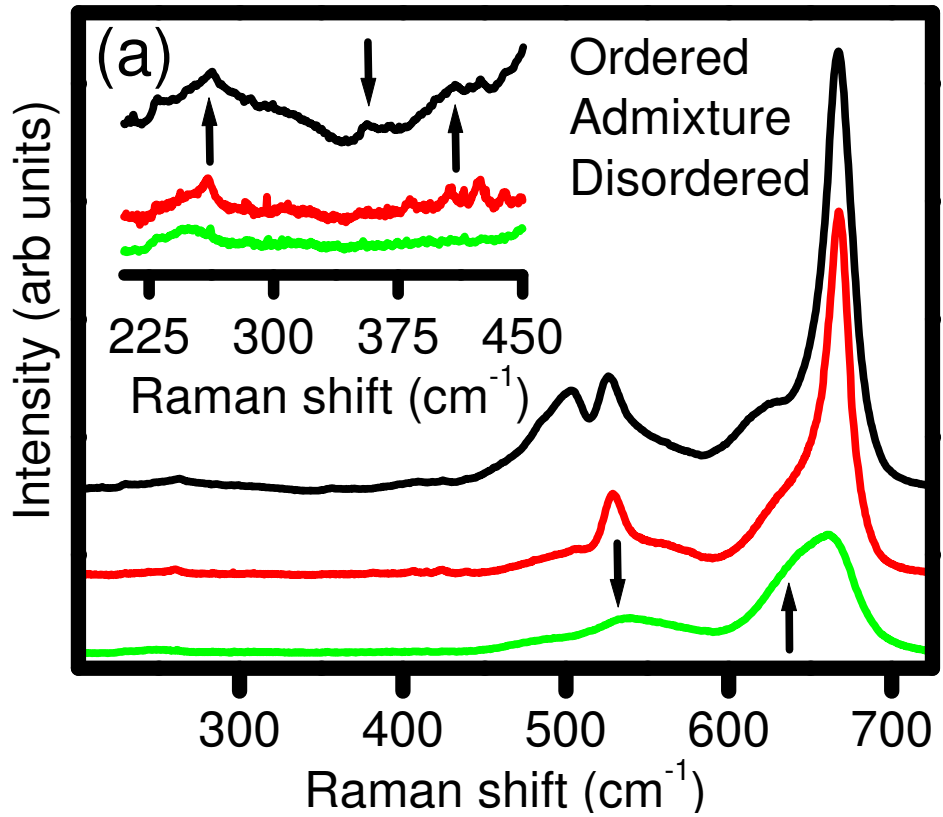
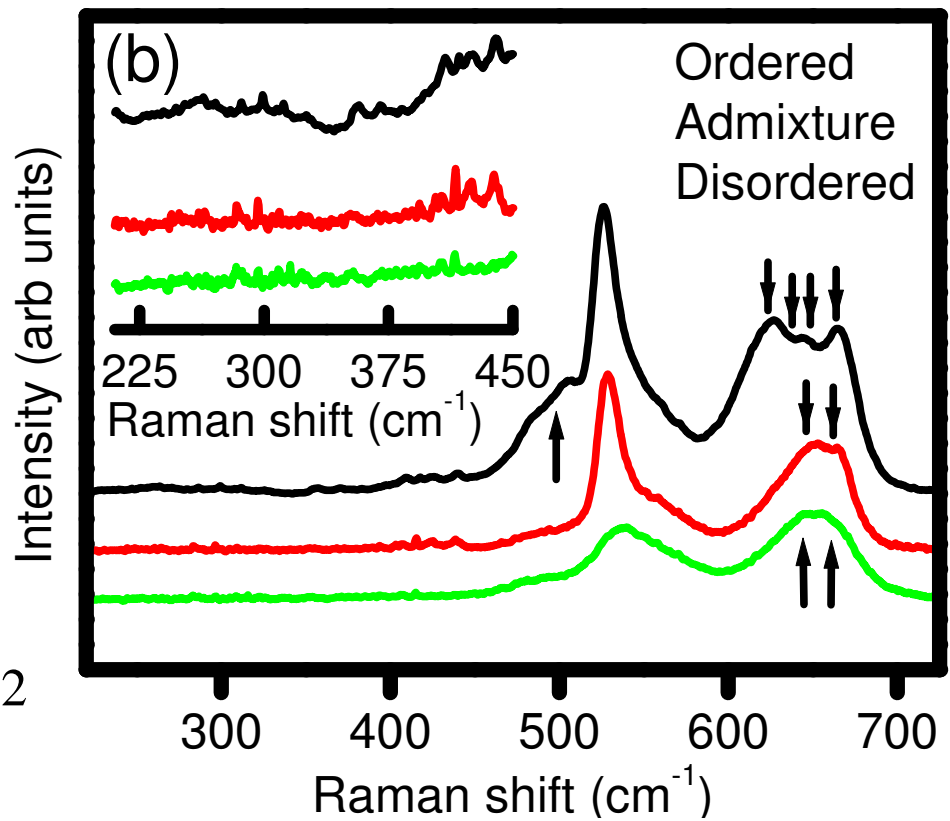

Figure 2

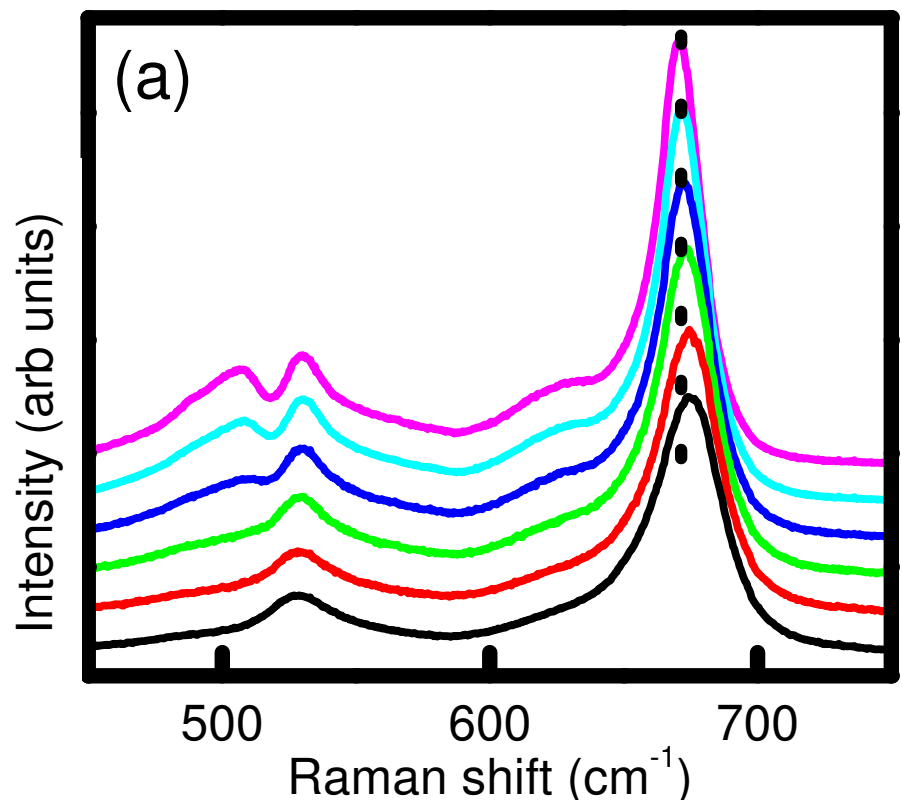

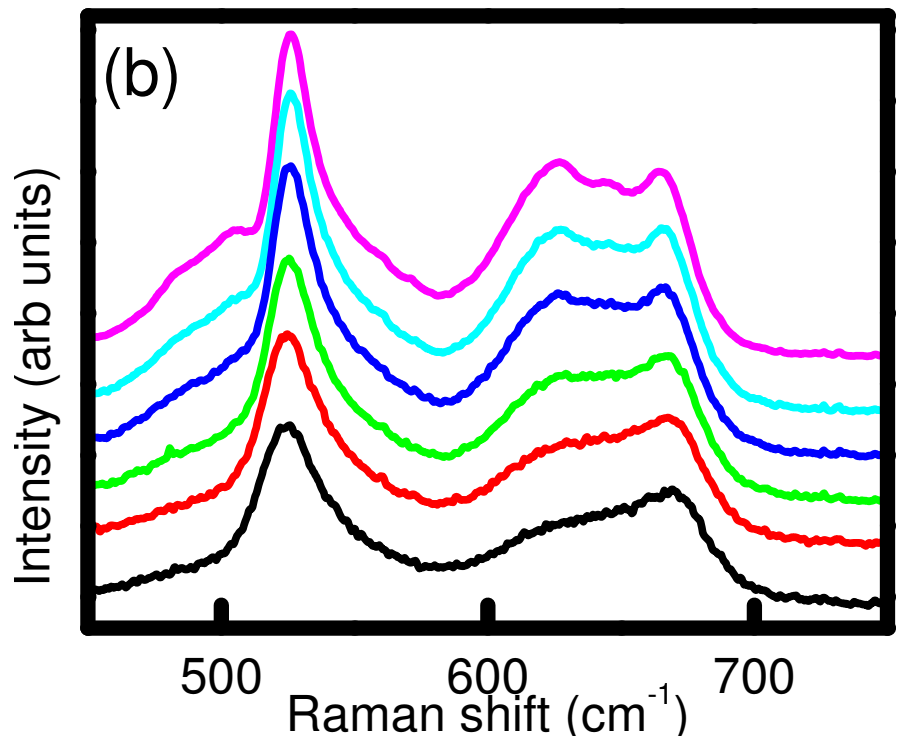

Figure 3

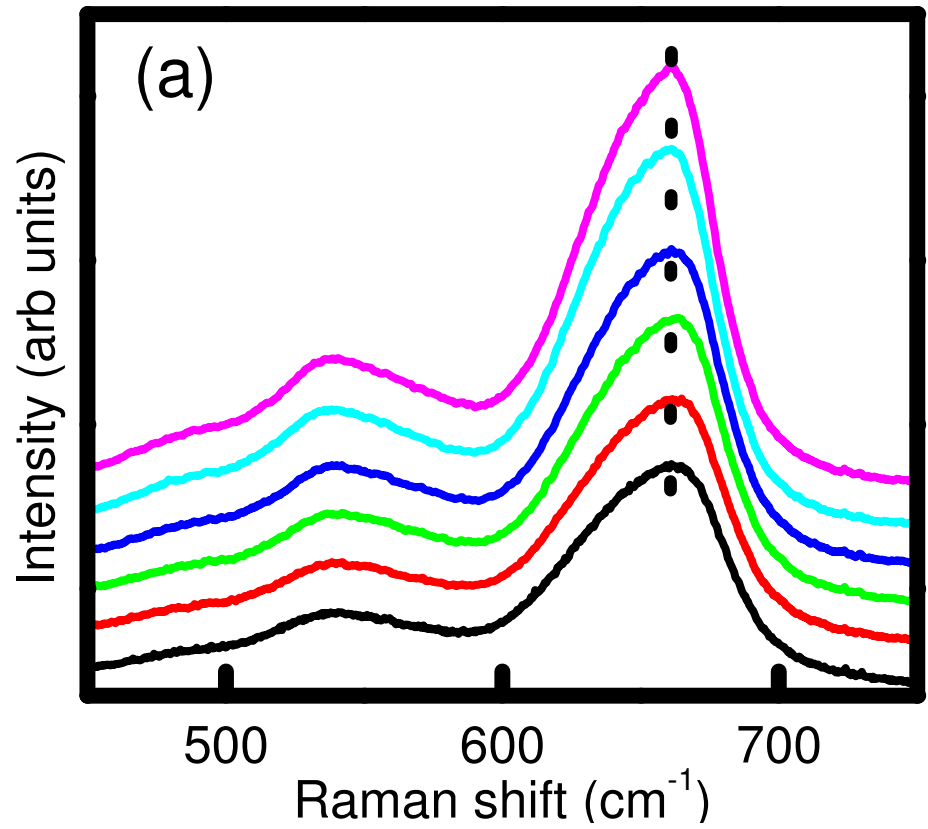
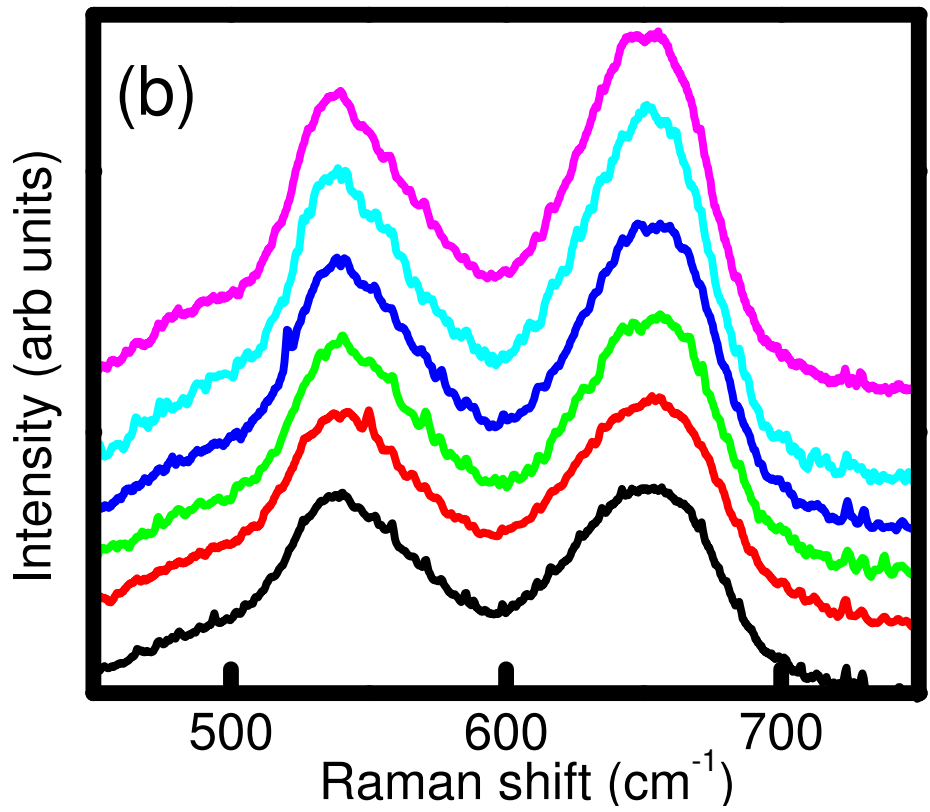

Figure 4

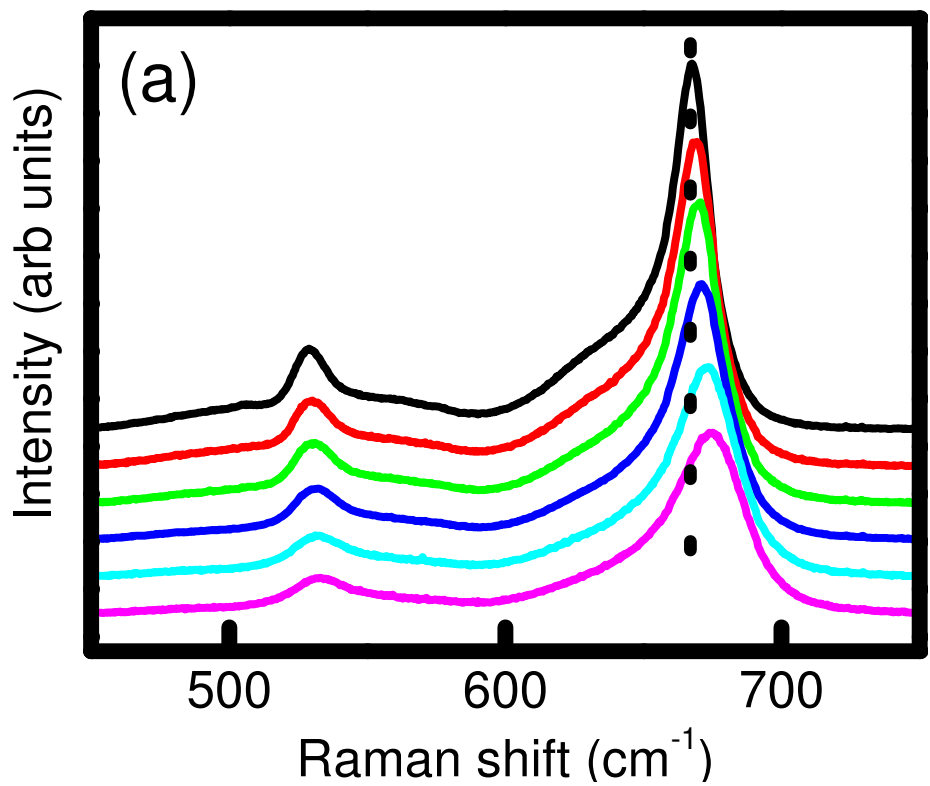

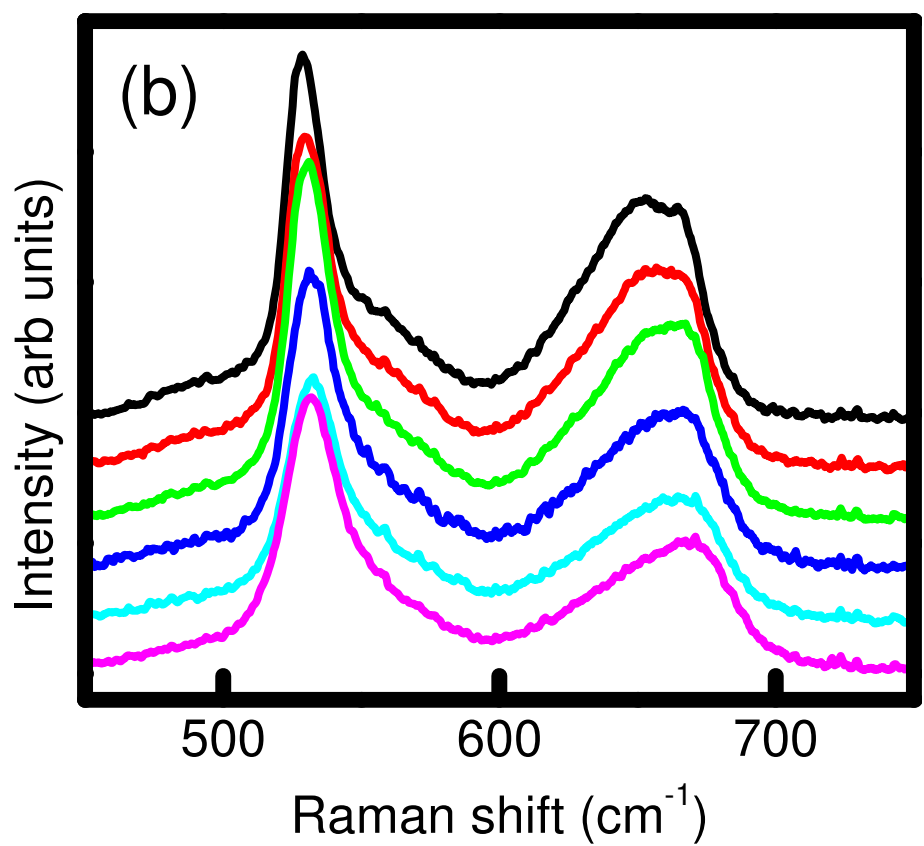

Figure 5

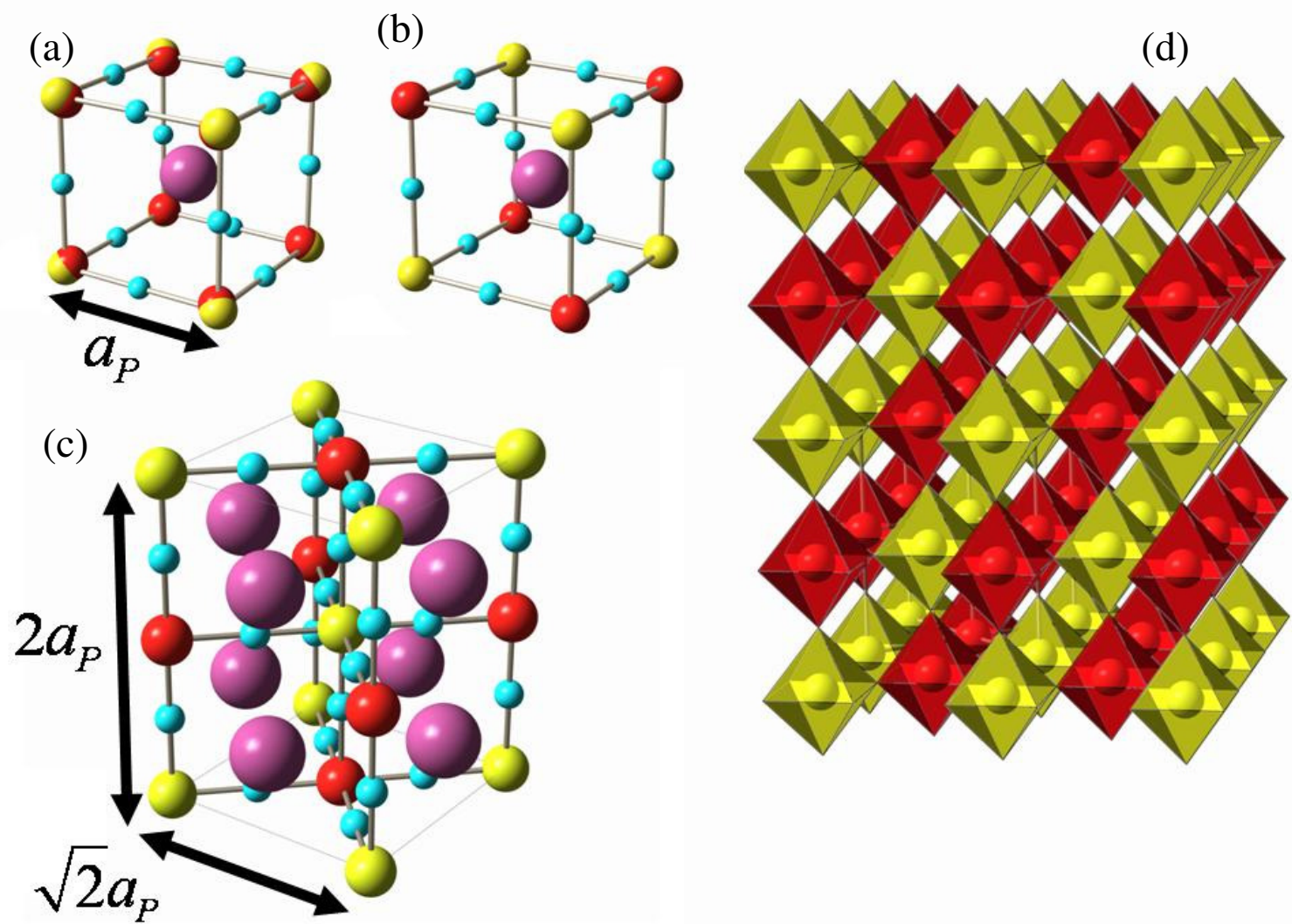

Figure 6

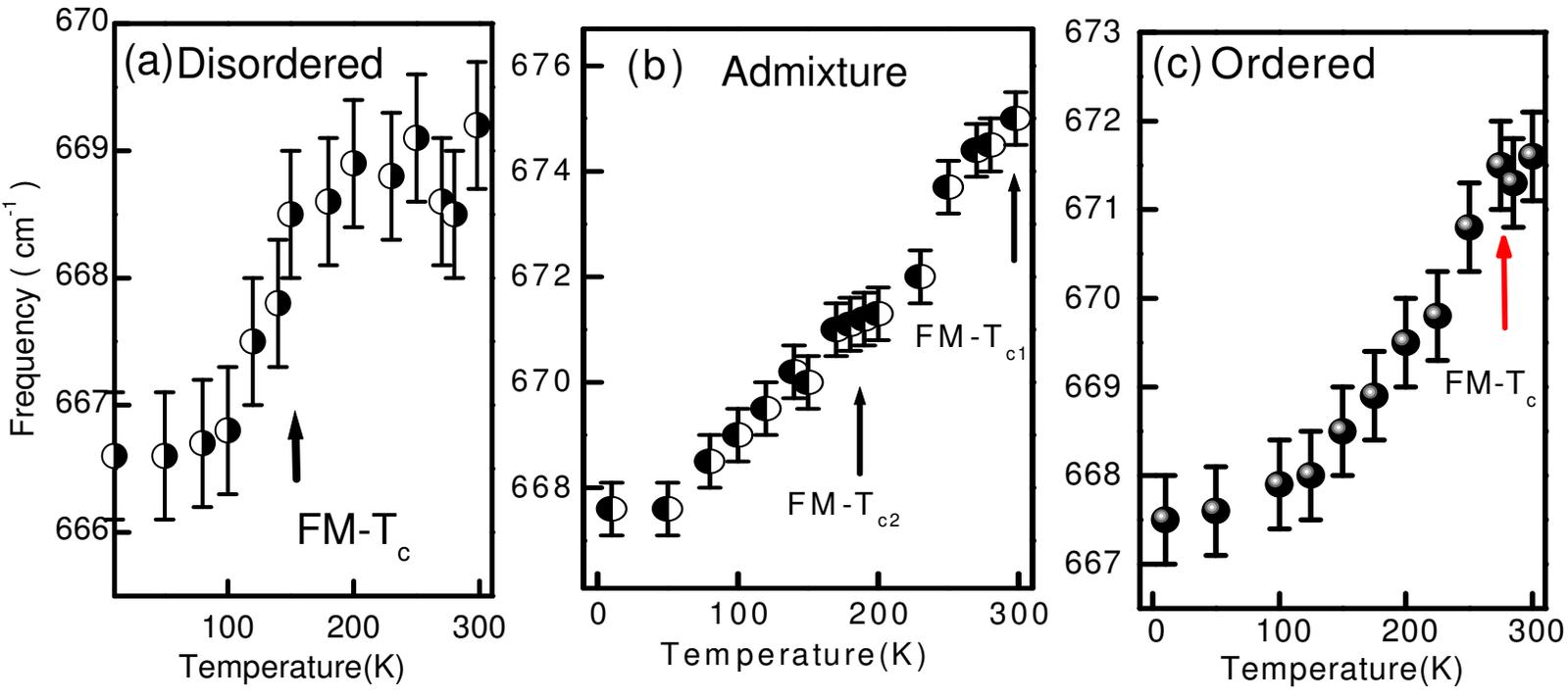

Figure 7